\documentclass[preprint,12pt]{elsarticle}

\usepackage{amssymb}

\usepackage{amsthm}
\usepackage{textcomp}

\journal{Optics Communications}

\begin{document}

\begin{frontmatter}



\title{Analysis of high-resolution spectra from a hybrid interferometric/dispersive spectrometer\tnoteref{label0}}

\tnotetext[label0]{\textcopyright~2015. This manuscript version is made available under the CC-BY-NC-ND 4.0 license http://creativecommons.org/licenses/by-nc-nd/4.0/}



\author[label1]{Phyllis Ko}
\address[label1]{The Pennsylvania State University,
University Park, PA 16802, USA}

\author[label2]{Jill R. Scott}
\address[label2]{Idaho National Laboratory, Idaho Falls, ID 83415, USA}

\author[label1]{Igor Jovanovic\corref{cor1}\fnref{label3}}
\ead{ijovanovic@psu.edu}

\begin{abstract}
To more fully take advantage of a low-cost, small footprint hybrid interferometric/dispersive spectrometer, a mathematical reconstruction technique was developed to accurately capture the high-resolution and relative peak intensities from complex spectral patterns. A Fabry-Perot etalon was coupled to a Czerny-Turner spectrometer, leading to increased spectral resolution by more than an order of magnitude without the commensurate increase in spectrometer size. Measurement of the industry standard Hg 313.1555/313.1844 nm doublet yielded a ratio of 0.682, which agreed well with an independent measurement and literature values. The doublet separation (29 pm) is similar to the U isotope shift (25 pm) at 424.437 nm that is of interest to monitoring nuclear nonproliferation activities. Additionally, the technique was applied to LIBS measurement of the mineral cinnabar (HgS) and resulted in a ratio of 0.682. This reconstruction method could enable significantly smaller, portable high-resolution instruments with isotopic specificity, benefiting a variety of spectroscopic applications. 
\end{abstract}


\end{frontmatter}


\section{Introduction}

Numerous scientific, medical, industrial, and verification applications demand high-resolution spectral measurements, which can be accomplished using Fabry-Perot (FP) interferometry, among other methods~\cite{VanDerWal1979,Zhou2012,Vogus2015}. Recently, FP interferometry has been adopted in pulsed measurements via laser-induced breakdown spectroscopy (LIBS)~\cite{Effenberger2012}. Spectra from laser-produced plasmas are commonly measured by use of angular dispersion, such as in Czerny-Turner (CT) or echelle spectrometers~\cite{Aragon2008}. In LIBS, high-resolution measurements are especially valuable for measurements of isotope ratios, where the typical isotopic shifts are on order a few to several tens of picometers. Currently, rapid and economical techniques with isotopic specificity are highly sought for nuclear safeguards and non-proliferation applications. Development of LIBS to detect and monitor facilities for illicit activities (\textit{e.g.}, isotope enrichment to undeclared levels)~\cite{Hanson2014,Ko2013} is critical to risk preparedness and treaty verification. However, high-resolution isotope measurements presently require very large (few-meter scale) CT spectrographs~\cite{Smith2002} or echelle spectrographs with complex instrument functions and limited dynamic range~\cite{Cremers2012}. More portable, simple, and cost-effective options are desirable in field applications. A practical alternative is an inexpensive, compact FP etalon coupled to a much lower resolution CT imaging spectrometer~\cite{Effenberger2012}. 

Accurate analysis of the 2-D intensity pattern measured by the instrument is important for interpreting profiles of spectral lines. A variety of algorithms have been developed to deconvolve the input spectrum from the interferometric measurement, but they rely on assumptions made about the shape (\textit{e.g.}, Gaussian, Lorentzian, Voigt functions) of the source spectrum~\cite{Larson1967,Koo1987,Naylor1991}. Other techniques include analysis in the Fourier domain, which also assumes a known source distribution~\cite{Cooper1971}. Direct least squares fitting of 2-D ring images from a single spectral line has been proposed, but is mainly aimed at calibrating and characterizing visible laser diodes~\cite{OHora2005}. Another alternative is data inversion~\cite{Zipoy1979,Heeg2007,Abbiss2008}, which can be effective in cases where the spectrum is unknown. Abbiss and Heeg measured spectra by varying the input angle of a tunable filter~\cite{Heeg2007}. While such scanning methods are not suitable for pulsed, imaging LIBS applications, similar mathematical principles were used in this case. The reconstruction method employed to interpret the high-resolution spectrum captured by a hybrid interferometric/dispersive spectrometer takes into account the intensity distribution of fringes observed in a 2-D image, and can be applied to spectra of a broadband source without presuming the shape of the input spectrum. In order to combine the benefits of 2-D algorithms with the speed necessary for rapid and high-resolution fringe measurement, this paper presents a simplified and robust fitting algorithm. The technique was also experimentally verified with an actual FP fringe imaging system.

The hybrid instrument, when used in conjunction with the reconstruction technique discussed here, represents an effective tool for measuring the high-resolution spectrum with accurate peak intensity ratios. Rapid, high-resolution spectral reconstruction capability from this type of hybrid measurement is ideal in certain applications of LIBS. The resolution of the hybrid interferometric-dispersive approach is at least an order of magnitude greater than the resolution of the CT spectrometer alone. This instrument can achieve the desired high resolution for measurement of the atomic isotopic shift of uranium, while reducing the size and cost of the instrument by approximately an order of magnitude.

\section{Instrument model and spectral analysis}

\begin{figure}[h!]
\centerline{\includegraphics[width=0.8\columnwidth]{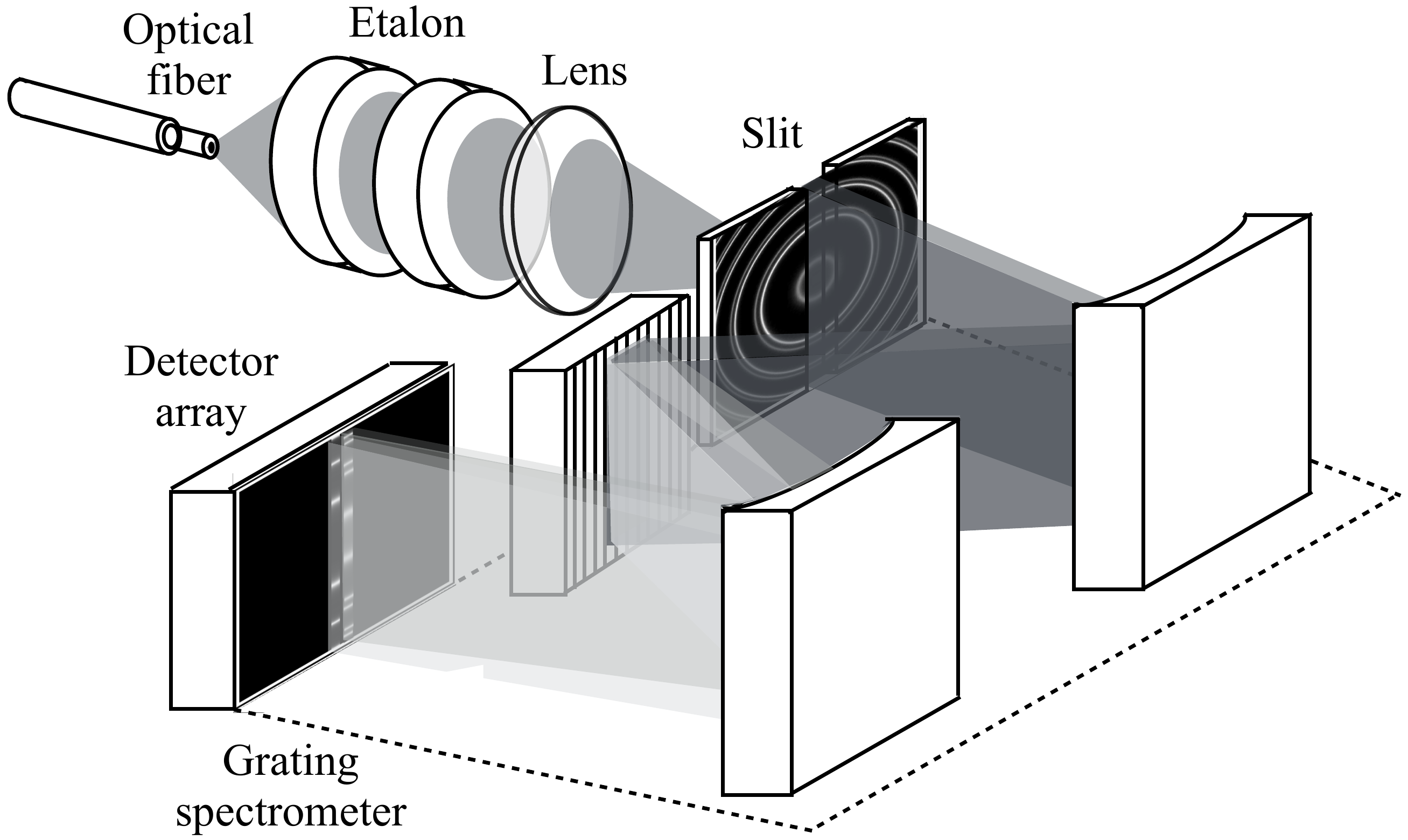}}
\caption{Schematic diagram of optical detection setup.}
\label{figure:Setup}
\end{figure}

The fringe pattern from an etalon exhibits angular and spectral dependence arising from constructive and destructive interference. A broad and complex spectrum incident on the FP etalon may produce a transmission pattern that cannot be uniquely reconstructed. This is due to degeneracy inherent in the interference pattern produced by the etalon, which exhibits a characteristic periodicity and wavelength-angle coupling, whereby an identical phase shift can be produced by different combinations of wavelength and transmission angle. This degeneracy can be broken by coupling the etalon to a dispersive spectrometer, which provides an additional angular dispersion, as described by Effenberger~\textit{et~al.}~\cite{Effenberger2012}. A diagram illustrating the approach to image sampling and dispersion is shown in Fig.~\ref{figure:Setup} and the image incident on the 2-D detector array is shown in Fig~\ref{figure:Normalize}(a). The high-resolution spectral information is encoded on the axis parallel to the spectrometer slit ({\it i.e.}, orthogonal to the dispersion plane of the spectrometer). The production of the interference ring pattern by the FP etalon can be described as
\begin{equation}
\label{equation:MatEq}
 B(\theta) = \int_0^\infty{T(\theta,\lambda)A(\theta,\lambda)d\lambda},
\end{equation}
where $A(\theta,\lambda)$ is the spectrally and angularly dependent intensity of the light incident on the etalon and $B(\theta)$ is the spectrally integrated intensity of the transmitted light. The FP etalon transmission function is described by the Airy equation,
\begin{equation}
\label{equation:Airy}
T(\theta, \lambda) = \left(1 + F\sin^2\left(
2 \pi d\,n/\lambda  \cos \theta
\right) \right)^{-1},
\end{equation}
where $F$ is the coefficient of finesse, given by $F=4R/\left(1-R\right)^2$ where $R$ is the etalon reflectivity, $d$ is the separation of the etalon mirrors, $n$ is the refractive index in the space between the mirrors, $\lambda$ is the wavelength, and $\theta$ is the angle measured with respect to the etalon axis. In a typical FP optical arrangement, the transmission pattern is measured in the focal plane of a lens (far field). The fringe pattern measured in this way is comprised of concentric rings of light corresponding to the angularly and spectrally dependent interference maxima arising from propagation through the FP etalon. The maxima appear when the condition $\cos \left(\theta\right)/\lambda = k \pi /\left(2 \pi d n\right)$ is met (where $k$ is an integer). Degeneracy arising from the coupling of angle and wavelength can be broken by introduction of angular dispersion, which leads to the measurement of
\begin{equation}
\label{equation:intpatt2D}
B(\theta,\lambda) = T(\theta,\lambda)A(\theta,\lambda).
\end{equation}
Here, $B(\theta,\lambda)$ represents the recorded 2-D interference pattern. The measurement of this pattern is typically limited by the available spectral resolution, which is dictated by the properties of the arrangement used to produce dispersion and the size of the detector pixels. At any wavelength $\lambda_0$, the quantity measured is
\begin{equation}
B(\theta,\lambda)|_{\lambda=\lambda_0} = \int_{\lambda_0-\epsilon/2}^{\lambda_0+\epsilon/2}T(\theta,\lambda)A(\theta,\lambda)d\lambda,
\end{equation}
where $\epsilon$ is the spectral resolution at wavelength $\lambda_0$. It is assumed for the purpose of this analysis that the resolution in the angular direction ($\theta$) is sufficient to accurately measure the shape of $B(\theta,\lambda)|_{\lambda=\lambda_0}$. It is further assumed that light incident on the etalon does not exhibit angular dependence of the spectrum, so that the angular dependence of $B(\theta,\lambda)$ arises solely from the nonuniform angular intensity profile and $B(\theta)=T(\theta,\lambda)A(\lambda)$. The light is incident on the etalon from the output of an optical fiber, and the Gaussian model chosen for the angular dependence of the intensity distribution is
\begin{equation}
\label{equation:fit}
B(\theta) = B_0 \exp \left(-(\theta-\theta_0)^2/w^2 \right) +C,
\end{equation}
where $B_0$ is the peak intensity of the Gaussian envelope, $\theta_0$ is the angle corresponding to the etalon axis, $C$ is a constant, and $w$ is the width of the distribution envelope~\cite{Scholl2004,Hirschberger2011}. The envelope function can be used to correct the measured FP pattern for its angular intensity dependence by fitting it to the Gaussian envelope $B(\theta)$.

This model of the measurement can be used for reconstructing the high-resolution features in the input spectrum directly from the measured pattern produced from the hybrid FP-dispersive instrument, without making additional assumptions regarding the shape of the spectrum. In a real measurement, the discretization and normalization (to the best-fit envelope function $A(\theta)$) of the measured transmission pattern in the neighborhood of the wavelength $\lambda_0$ yields the vector
\begin{equation}
\mathbf{B}_\theta = \mathbf{T}_{\theta,\lambda} \mathbf{A}_{\lambda}.
\end{equation}
Here, $\mathbf{T}_{\theta,\lambda}$ is the transmission matrix constructed by discretization of Eq.~(\ref{equation:Airy}), and $\mathbf{A}_{\lambda}$ is the vector representing the spectrum in the vicinity of the wavelength $\lambda_0$. The high-resolution spectrum can then be reconstructed as
\begin{equation}
\label{equation:inverse}
\mathbf{A}_\lambda = \mathbf{T}^{-1}_{\theta,\lambda} \mathbf{B}_{\theta},
\end{equation}
which is, in practice, reduced to finding the least-squares solution to this inverse problem. 

\section{Experimental results and discussion}

The reconstruction method was applied to data collected from a hybrid FP instrument that used a 550 mm focal length imaging spectrometer (Horiba Jobin Yvon) equipped with a 1800~mm$^{-1}$ grating and a 1024$\times$1024 pixel array \mbox{iStar} ICCD (Andor) composed of 13~\textmu m pixels. Coupled to the entrance of the spectrometer was a FP etalon (SLS Optics). Light was transported to the etalon by a multimode 400~\textmu m core diameter optical fiber (Ocean Optics). Fiber-optic connector, etalon, and lens were affixed in a cage setup, with multi-axis adjustments to enable accurate alignment of the circular transmission pattern to the spectrometer slit. A 25~mm diameter lens with a focal length of 100~mm was used to focus the output of the etalon onto the entrance slit of the spectrometer. The light source used was a Hg(Ar) calibration pen lamp (Oriel).

\begin{figure}[h!]
\centerline{\includegraphics[width=1\columnwidth]{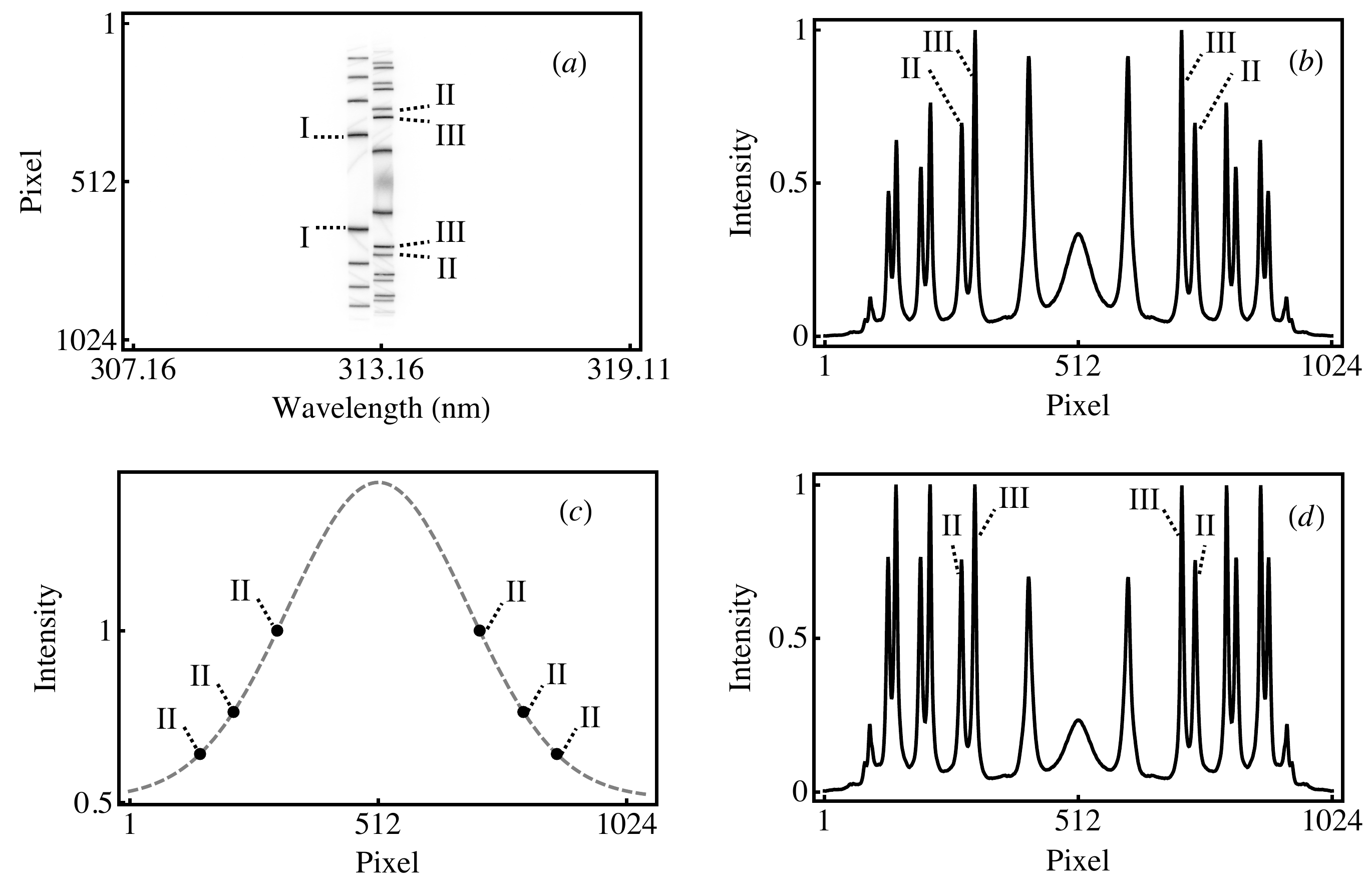}}
\caption{(a) Measured pattern corresponding to the (I) 312.5674~nm line and (II) 313.1555/(III) 313.1844~nm doublet; (b) 1-D integrated profile of doublet; (c) intensity envelope function for (II) 313.155 nm (dots -- experimental data; dashed line -- fit to Eq.~(\ref{equation:fit})); (d) normalized 1-D profile.}
\label{figure:Normalize}
\end{figure}

The FP etalon was coupled to the entrance slit of the spectrometer to produce fringes, as shown in Fig.~\ref{figure:Normalize}(a). The section of the spectrum corresponding to the 313.1555/313.1844~nm doublet was summed along the horizontal axis to obtain the 1-D profile of the transmitted intensity, shown in Fig.~\ref{figure:Normalize}(b). The symmetric pattern was collapsed to summing about the etalon axis to improve the signal-to-noise ratio (SNR). A Gaussian envelope function was fit to the fringes exhibiting a characteristic double-peaked shape to obtain the $A(\theta)$ correction, as shown in Fig.~\ref{figure:Normalize}(c). The normalized transmitted intensity pattern is shown in Fig.~\ref{figure:Normalize}(d).

\begin{figure}[h!]
\centerline{\includegraphics[width=0.8\columnwidth]{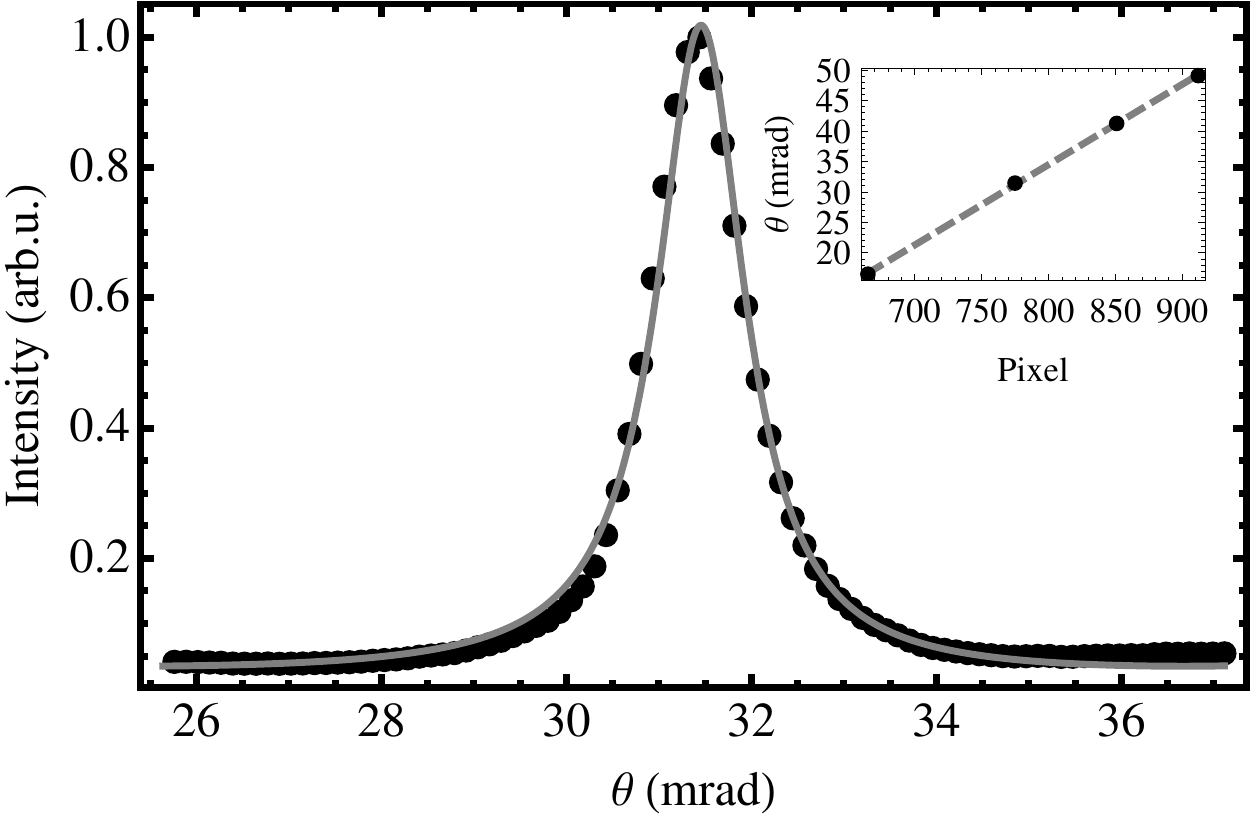}}
\caption{Best-fit (solid line) of the measured 312.5674~nm first-order interference peak by Eq.~(\ref{equation:Airy}) yields the reflectivity of the etalon ($R=0.73$). Experimentally measured points are shown as black markers. Inset shows calibration of the angular ($\theta$) axis using the location of four 312.5674~nm constructive interference peaks (black markers), The linear fit is shown as a dashed line.}
\label{figure:FitR}
\end{figure}

It is necessary to calibrate the angular axis before formulating the instrument function. This calibration was performed using the neighboring 312.5674~nm line (Fig.~\ref{figure:Normalize}(a)). The angular position of highest intensity (peak) in the fringe profile was used to perform the calibration in Fig.~\ref{figure:FitR}. Because the paraxial approximation is valid, the calibration can be described by a linear function, also shown in Fig.~\ref{figure:FitR}.
The region of the fringes selected for this fit did not include the area near the optical axis or the outermost fringes, which may exhibit distortions due to their low intensity and possible optical aberrations.

Next, the instrument function for the 313~nm doublet was determined. The ranges of $\theta$ and $\lambda$ used to construct the instrument function were 20 mrad $<\theta<$ 32 mrad and 313.106~nm $<\lambda<$ 313.234~nm, respectively. The etalon had air-spaced ($n=1$) mirrors separated by a distance of $d=0.44$~mm. The etalon reflectivity ($R=0.73$) was determined by fitting the measured intensity profile to the known FP etalon transmission function, Eq. (\ref{equation:Airy}), as shown in Fig.~\ref{figure:FitR}. 

Interpolation between points and increased sampling of the experimental spectrum was performed prior to reconstruction of the source spectrum, $\mathbf{A}_\lambda$. The least-squares solution to the system of linear equations associated with Eq.~(\ref{equation:inverse}) could have been used to reconstruct $\mathbf{A}_\lambda$. However, Eq.~(\ref{equation:inverse}) is an example of a problem where $\mathbf{T}_{\theta, \lambda}$ is an ill-conditioned matrix~\cite{Abbiss2008}. Hence, the pseudoinverse $\mathbf{T}_{\theta, \lambda}^{+}$ of the matrix was used to calculate a least-squares solution to the system of equations using
\begin{equation}
\mathbf{T}_{\theta,\lambda}^{+} \mathbf{B}_\theta = \mathbf{A}_\lambda.
\end{equation}
The pseudoinverse solution was implemented in Mathematica (Wolfram Research) using the function \texttt{pseudoinverse}. Because of the ill-condition of the matrix $\mathbf{T}_{\theta,\lambda}$, it was necessary to specify the tolerance parameter. The singular values matrix are the non-negative square roots of the eigenvalues of $\mathbf{T}_{\theta,\lambda}^* \mathbf{T}_{\theta,\lambda}$, where $\mathbf{T}_{\theta,\lambda}^*$ is the conjugated transposed matrix. Singular values smaller than the product of the tolerance and the largest singular value of $\mathbf{T}_{\theta,\lambda}$ were omitted~\cite{Golub1970}. In practice, the choice of tolerance controls the influence of smaller singular values in the composite error of the reconstruction, and is a trade-off between reducing round-off error at the expense of a larger residual. A tolerance of 0.1 was used throughout the analysis to yield the optimal approximation to the solution.

Minimization of the residual error of the reconstruction was the next step. The widths of Hg lines from a low-pressure discharge calibration lamp are approximately 1~pm~\cite{Cremers2012}. The width of the reconstructed spectrum is dictated primarily by the reflectivity $R$. Because the widths are narrow compared to the instrument function, an accurate reconstruction assuming this experimentally determined reflectivity was not possible. By assuming a broader source spectrum, it was possible to minimize residual error in the spectral reconstruction. A reflectivity value of $R_\alpha = \alpha R$, where $\alpha$ is a multiplicative constant of $R$ that minimizes the residual error, $\parallel \mathbf{B}_{\theta} - \mathbf{T}_{\lambda, \theta} \mathbf{A}_{\lambda} \parallel$ was assumed. An optimal value $\alpha = 1.16$, obtained by scanning a range of values 1.00$<\alpha<$1.30, was utilized to minimize the residual error of the reconstruction.

\begin{figure}[h!]
\centerline{\includegraphics[width=0.8\columnwidth]{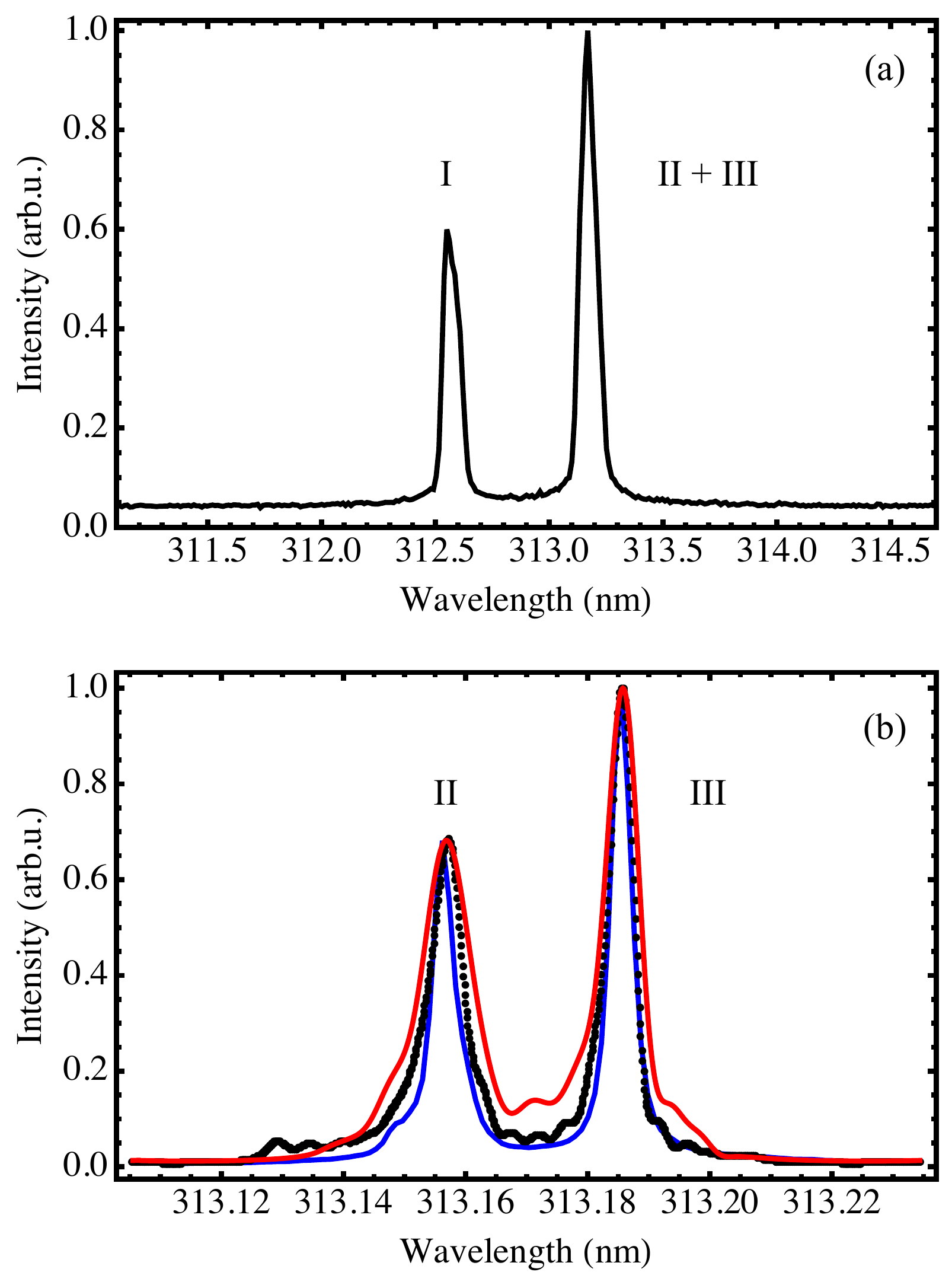}}
\caption{ (Color online) (a) The (I) 312.5674~nm line and (II) 313.1555/(III) 313.1844~nm doublet measured with the CT spectrometer alone; the doublet (II+III) was not resolved. (b) The spectrum of the (II) 313.1555/(III) 313.1844~nm doublet (black, dotted) reconstructed from the hybrid FP/CT measurement compared to DEMON measurement (blue, solid). LIBS of HgS reconstructed from a hybrid FP/CT measurement is also shown (red, dashed).}
\label{figure:Reconstruct}
\end{figure}

A spectrum of Hg measured using the CT spectrometer without the FP etalon is shown in Fig.~4(a) for comparison. In this case the resolution ($\sim$70~pm) was not sufficient for resolving the 313~nm doublet. The high-resolution reconstruction provided by the hybrid instrument is shown in Fig.~\ref{figure:Reconstruct}(b) and summarized in Table~1. The resolution of the 313.1844~nm line was $\sim$4~pm, which exceeds the resolution of the CT spectrometer by more than one order of magnitude. These results demonstrate that, while the resolution results from the design of the entire hybrid instrument, the resolution enhancement is a consequence of the introduction of the FP etalon. An accurate reconstruction method is critical to fully realize the utility of a hybrid instrument. For comparison, the 313.1844~nm line in Fig.~\ref{figure:Normalize}(d) has a resolution of $\sim$8 pm using more traditional methods for converting the fringe pattern to a spectrum. Additionally, Effenberger~\textit{et~al.}~estimated the resolution to be $\sim$11~pm from the measured peak separation in spectra acquired with an Hg(Ar) lamp coupled into a FP attached to a 500~mm CT spectrometer with a 512$\times$512, 24~\textmu m/pixel detector array.

Correcting for the angular drop in intensity of the measured FP pattern was critical for peak ratio determination; the reconstructed peaks of non-normalized data yielded a ratio of 0.629, which is lower due to the decreasing intensity as a function of distance from the interferogram center. Validation of the reconstructed hybrid instrument spectrum was performed by comparing the results with those acquired with a high-resolution echelle spectrometer (DEMON, Lasertechnik Berlin) using the same Hg(Ar) lamp. The DEMON results are also provided in Fig.~\ref{figure:Reconstruct} and Table~\ref{table:Ratios}, along with literature values~\cite{Reader1996}.

Averages reported in Table~\ref{table:Ratios} are based on 10, 5, and 20 spectra for sources 1, 2, and 3, respectively. The reconstructed peak wavelengths were 313.156~nm and 313.185~nm for the Hg doublet, and agree well with actual values. While the resolution is not as high for the hybrid instrument, the spectra are similar (Fig.~\ref{figure:Reconstruct}(b)) and give a very similar ratio accuracy (Table~\ref{table:Ratios}).
\begin{table}[h!]
  \caption{Peak ratios of the 313~nm doublet of a Hg(Ar) calibration lamp measured by different sources. The same lamp was measured by method 1 and 2.}
  \begin{center}
    \begin{tabular}{ c c c | c c c | c c c | c c c }
    & Source & & & Ratio & & & Uncert.(\%) & & & $\lambda/\Delta\lambda$ & \\
    \hline
    1. & Hybrid device & & & 0.682 & & & 1.8 & & & $\sim$37,000 & \\
    2. & DEMON & & & 0.681 & & & 2.4 & & & $\sim$75,000 & \\
    3. & Ref.~\cite{Reader1996} & & & 0.687 & & & 7.2 & & & N/A & \\
    \hline
    \end{tabular}
  \end{center}
  \label{table:Ratios}
\end{table}

To further explore the utility of this spectral reconstruction method, it was applied to the analysis of LIBS data from the mineral cinnabar (HgS) acquired with a different hybrid spectrometer~\cite{Effenberger2012}. In this case, plasma was generated by 1064 nm laser pulses at 25 mJ per pulse. The spectrum was collected from the accumulation of 600 laser pulses at a rate of 10 Hz and under 10 Torr He atmosphere, which is the optimum pressure for the Hg 313~nm doublet~\cite{Effenberger2012}. The reconstructed doublet, shown in Fig.~\ref{figure:Reconstruct}(b), exhibits a ratio of 0.682 and is in good agreement with values reported in Table~\ref{table:Ratios}. Peak broadening in the LIBS spectrum is due to plasma dynamics~\cite{Scott2014}.

\section{Conclusion}

This spectral reconstruction approach for hybrid FP/dispersive spectrometers demonstrates that simple, low-cost instruments with high resolution and accurate relative peak intensities can be realized. Accurate wavelength and ratio determination of Hg lines using this instrument has been presented, where the accuracy depends on proper normalization and calibration of the measured interferogram. Because only a fraction of light is coupled to the CT spectrometer, a fraction of the signal is lost in this method. Summation of signal from different orders produced by the FP etalon may be able to increase the SNR, but was not implemented in the reconstruction discussed here. Experimentally, a higher repetition rate laser may be used to speed collection time and further increase the SNR. While such instruments would be invaluable for LIBS measurements requiring accurate isotope ratio measurements, the technique could enable practical instruments for a variety of high resolution field applications.\\

\newpage

The authors would like to thank Andrew J.~Effenberger, Jr.~and Elizabeth J.~Judge for their assistance. Research was performed under appointment to the Nuclear Nonproliferation International Safeguards Graduate Fellowship Program sponsored by the National Nuclear Security Administration's Next Generation Safeguards Initiative (NGSI). Material is based upon work supported by the U.S. Department of Homeland Security under Grant Award Number, 2012-DN-130-NF0001-02 and in part funded by the Consortium for Verification Technology under Department of Energy National Nuclear Security Administration award number DE-NA0002534. The views and conclusions contained in this document are those of the authors and should not be interpreted as necessarily representing the official policies, either expressed or implied, of the U.S. Department of Homeland Security. Research was also sponsored by the U.S. Department of Energy under DOE Idaho Operations Office Contract DE-AC07-05ID14517.





\newpage

\newpage

\newpage

\newpage

\end{document}